\documentclass[journal,draftcls,onecolumn,12pt,twoside]{IEEEtranTCOM}

\normalsize

\usepackage{amsmath}
\usepackage{graphicx}
\usepackage{amssymb}
\usepackage{amsthm}
\usepackage{cite}
\usepackage{mathtools}
\usepackage{bbold}
\usepackage{steinmetz}
\usepackage{stmaryrd}
\usepackage{bbm, dsfont}
\usepackage{verbatim}
\usepackage[latin1]{inputenc}
\usepackage{tikz}
\usetikzlibrary{shapes,arrows}
\usepackage{caption}
\usepackage{enumerate}
\usepackage{color}
\usepackage{subfig}
\usepackage{pgfplots}
\usetikzlibrary{calc}

\usepackage{lipsum}
\usepackage[variablett]{lmodern}
\usepackage{amsbsy}
\usepackage{bm}
\usetikzlibrary{positioning}%
\usetikzlibrary{shapes,arrows}

\pgfplotsset{
	every tick label/.append style={scale=1},
	every axis/.append style={
	}
}

\newcommand{\ai}{a}

\newcommand{\azexp}[1]{\tilde{\underline{\mathbf{a}}}_{#1}}
\newcommand{\azexpt}[2]{\tilde{{\mathbf{a}}}_{#1,#2}}

\newcommand{\lone}{L_K^{(1)}}
\newcommand{\ltwo}{{L}_K^{(2)}}
\newcommand{\lthree}{{L}_K^{(3)}}
\newcommand{\ltwoinf}{\tilde{L}_K^{(2)}}
\newcommand{\lthreeinf}{\tilde{L}_K^{(3)}}
\newcommand{\daf}{\mathbf{w}}
\newcommand{\di}{\mathrm{d}}

\DeclareMathOperator{\Tr}{Tr}					               %
\newcommand{\red}{\mathcal{R}}                                               %
\newcommand{\sn}{\sigma_n}

\newcommand{\Cone}{C_1}                                               %
\newcommand{\Ctwo}{C_2}                                               %
\newcommand{\ima}{\mathcal{I}}                                                %
\newcommand{\cpct}{\mathcal{C}}                                                %
\newcommand{\mf}{\mathbf}                                                %
\newcommand{\R}{\mathsf{R}}                                                                  %
\newcommand{\UB}{D}                     %
\newcommand{\UBtw}{G}                     %
\newcommand{\alf}[2]{\alpha_{#1,#2}}                     %
\newcommand{\nzn}[1]{\underline{\mathbf{n}}_{#1}}                     %
\newcommand{\az}[1]{\underline{\mathbf{a}}_{#1}}                     %
\newcommand{\anf}[1]{\underline{\mathbf{\tilde{a}}}^{\mathrm{nl}}_{#1}}                     %
\newcommand{\an}[1]{\underline{\mathbf{a}}^{\mathrm{nl}}_{#1}}                     %
\newcommand{\al}[1]{\underline{\mathbf{a}}^{\mathrm{li}}_{#1}}                     %

\newcommand{\azt}[2]{{\mathbf{a}}_{#1,#2}}                     %
\newcommand{\ant}[2]{{\mathbf{a}}^{\mathrm{nl}}_{#1,#2}}                     %
\newcommand{\antf}[2]{{\mathbf{\tilde{a}}}^{\mathrm{nl}}_{#1,#2}}                     %

\newcommand{\alt}[2]{{\mathbf{a}}^{\mathrm{li}}_{#1,#2}}                     %

\newcommand{\ants}[2]{\bigl({\mathbf{a}}^{\mathrm{nl}}_{#1,#2}\bigr)^*}                     %

\newcommand{\dll}[1]{{d}_{#1}}

\newcommand{\zf}{Z}                                                                   %

\newcommand{\nt}{L}                                                                    %
\newcommand{\lo}{{\mathsf{U}^{\mathrm{li}}}}                                                                %
\newcommand{\nzt}[2]{{\mathbf{n}}_{#1,#2}}                       %
\newcommand{\tnp}{N_{ASE}}                                                       %
\newcommand{\dz}{\Delta_z}                                                          %
\newcommand{\dt}{\Delta_t}						            %
\newcommand{\ff}{\mathsf{F}}									    %
\newcommand{\abx}[1]{\llbracket #1 \rrbracket  ^2}                   %
\newcommand{\abz}[1]{\llbracket #1 \rrbracket ^2}                                      %

\newcommand{\expe}{\mathrm{E}}

\newcommand{\ea}{\mathrm{E}_{\underline{\mathbf{a}}_{0}}}

\newcommand{\absol}[1]{|#1|}
\newcommand{\absl}[1]{|#1|}

\newcommand{\ang}[1]{\phase{#1}}

\newcommand{\CC}{\mathbb{C}}
\newcommand{\xx}{\underline{\mathbf{x}}}
\newcommand{\ww}{\underline{\mathbf{w}}}
\newcommand{\yy}{\underline{\mathbf{y}}}

\newcommand{\dllf}{{f^{\mathrm{li}}}}

\newcommand{\np}{\mathbf{\underline{n}}_p}
\newcommand{\npt}[1]{\mathbf{n}_{p,#1}}
\newcommand{\pnexp}{\tilde{P}_n}

\newcommand{\ul}[1]{\underline{#1}}                  
                  
\newcommand{\be}{\begin{equation}}
\newcommand{\ee}{\end{equation}}
\newcommand{\bs}{\begin{split}}
	\newcommand{\es}{\end{split}}
\newcommand{\bes}{\begin{equation}\begin{split}}
\newtheorem{theorem}{Theorem}

\newtheorem{lemma}{Lemma}
\newcommand{\bal}{\begin{align}}

\newcommand{\eal}{\end{align}}

\newcommand{\rlp}[1]{\left(#1\right)}
\newcommand{\rla}[1]{\left\{#1\right\}}
\newcommand{\rlb}[1]{\left[#1\right]}
\newcommand{\rlpo}[1]{\lefto(#1\right)}
\newcommand{\rlao}[1]{\lefto\{#1\right\}}
\newcommand{\rlbo}[1]{\lefto[#1\right]}
\newcommand{\po}{P}
\newcommand{\pn}{P_n}

\newcommand{\lefto}{\mathopen{}\left}
\newcommand{\vecnorm}[1]{\lVert#1\rVert}		%
\newcommand{\jpg}{\mathcal{CN}}			%
\newcommand{\distas}{\sim}					%
\newcommand{\given}{\,\vert\,}				%

\begin{document}
\title{ Rates Achievable on a Fiber-Optical Split-Step Fourier  Channel}

\author{Kamran~Keykhosravi,~\IEEEmembership{Student~Member,~IEEE,}
        Erik~Agrell,~\IEEEmembership{Senior~Member,~IEEE,}
        Giuseppe~Durisi,~\IEEEmembership{Senior~Member,~IEEE}%
\thanks{ This work
	was supported by the Swedish Research Council (VR) under Grant 2013-5271. The simulations were executed  on resources provided by the
	Swedish National Infrastructure for Computing (SNIC) at C3SE.  This work was presented in part at the Munich Workshop on Information Theory of Optical Fiber, 2015 (not published as a paper). }
\thanks{The authors are with the Department of Signals and Systems,
	Chalmers University of Technology, Gothenburg 41296, Sweden (e-mail:
	kamrank@chalmers.se; durisi@chalmers.se; agrell@chalmers.se).}%
}

\maketitle

\begin{abstract}
	    A  lower bound on the capacity of  the split-step Fourier channel is derived. The channel under study is a concatenation  of smaller segments,  within which three operations are performed on the signal, namely, nonlinearity, linearity, and noise addition.
	    Simulation results indicate that  for a fixed number of segments, our lower bound  saturates in the high-power regime and that the larger the number of segments is, the higher is the saturation point.
	    We also obtain an alternative lower bound, which is less tight but has a simple closed-form expression. This bound allows us to conclude that the saturation point grows unbounded with the number of segments. 
	    Specifically,
	     it grows  as $c+(1/2)\log\rlpo{K}$, where $K$ is the number of segments and $c$ is a constant. 
	    The connection between our channel model and the  nonlinear Schr\"odinger equation is  discussed.  
\end{abstract}

\IEEEpeerreviewmaketitle

\section{Introduction}\label{sec:intro}
\IEEEPARstart{F}{inding}  the capacity of the nonlinear and dispersive optical channel is a formidable task, so much so that not only the capacity has not been established, but also  a large gap between the known upper and lower bounds exists. 
While all known lower bounds either saturate %
 or fall to zero
 in the high-power regime, the only available upper bound \cite{Kramer_2015_itw,Yousefi_2015_cwit} grows logarithmic with the power, i.e., it behaves as $\log\lefto(1+\mathrm{SNR}\right) $, where $\mathrm{SNR}$ is the signal-to-noise ratio.
Neglecting  dispersion, the channel capacity can be calculated \cite{yousefi_2011_tinf,turitsyn_2003_prl}. In this case, its asymptotic behavior  in the high-power regime is $(1/2)\log\lefto(\mathrm{SNR}\right)-1/2$. 

Many lower bounds have been proposed on the capacity of fiber-optical channels (see for example \cite{mitra2001nonlinear,Ellis_2010_jlt,Meccozzi_2012_jlt,irukulapati2016tighter,secondini_2013_jlt}), most of which fall to zero at high powers. 
Consequently, it was widely believed that the capacity would diminish at high powers. 
Recent works disprove this belief  \cite{agrell2015conditions,agrell_2014_jlt}. However, to the best of our knowledge, no lower bound has been established as yet that grows unbounded with power.

The nonlinear Schr\"odinger equation (NLSE) models the fiber-optical channel excellently. 
However, it is not suitable for information theory analyses since  its input and output are continuous-time waveforms. 
The split-step Fourier (SSF) method is a standard method to simulate the NLSE and has been validated by many experiments. 
The SSF method approximates the NLSE by discretizing it in time and space (by splitting the fiber channel into multiple segments).  
Thus, its input and output  are complex vectors. 
Moreover, the output vector can be obtained by recursive computations  over the many channel segments. 
This method has been used in \cite{Kramer_2015_itw} to establish an upper bound on the capacity of the fiber-optical channel.  

The accuracy of the SSF method depends on the step size in the spatial domain as well as the sampling interval in the temporal domain. 
When the number of segments goes to infinity, the SSF method approximates the NLSE accurately \cite[Sec.~4.2.1]{agrawal_2007_nfo}. 
The error caused by using  a finite  number of segments depends on the input power: to maintain a desired accuracy level as the input power grows,  the number of segments needs to increase (or, equivalently, the step size needs to decrease) with the input power.
 To the best of our knowledge, no  closed-form expression is available for the  SSF method error as a function of the input power and the step size.
  However, results based on simulations and on approximations of the error are available (see for example \cite{zhang_symmetrize}).
  
   Because of  the nonlinearity, the bandwidth of the input signal  broadens as it propagates along the fiber.
In practice, this effect is taken into account in the SSF model by oversampling, i.e., by sampling faster than the Nyquist rate. In this paper, we  ignore the effects of spectrum broadening, which is left to future studies, and consider an SSF model with sampling performed at the Nyquist rate.

\textit{Contributions:} We derive a lower bound on the capacity of the SSF fiber-optical channel with $K$ segments, using as input vector distribution a multivariate Gaussian with independent and identically distributed (i.i.d.) components. We present a lower bound that can be evaluated by calculating through Monte Carlo simulations the  expectation of a function of  input and noise vectors.  The simulation results indicate that the bound  saturates at high powers and that the saturation point increases with the number of  segments.   We further lower-bound the aforementioned  bound by a  closed-form expression, which  reveals that the saturation point increases by 0.5 bit whenever the number of segments is doubled. This unbounded increase of our capacity lower bound casts doubt on the existence of an optimal input power, going beyond which is  idle or even detrimental to the optical system performance.

The outline of this paper is as follows. 
In Section~\ref{s2}, we describe the continuous-time and the discrete-time channel models. 
In Section~\ref{s3}, we state our simulation-based as well as closed-form lower bounds. The simulation results are provided in Section~\ref{sec:simulation}.
The first Appendix contains some preliminary results, which come into use in the subsequent Appendices, where  our theorems are proved.

\paragraph*{Notation}
We use  boldface letters to denote random quantities. 
Vectors, which are columns by default, are identified by underlined letters, whereas matrices are denoted by upper-case \emph{sans-serif} letters  (e.g., $\mathsf{A}$).
The identity matrix of size $L\times L$ is denoted by $\mathsf{I}_L$.
The $i$th element of a vector is indicated by the subscript $i$.
 For a complex number $x$, we  denote its real part, imaginary part, absolute value, and phase by  $\red\lefto( x\right)$, $\ima\lefto( x\right)$, $\absol{ x}$, and $\ang{ x}$, respectively. 
The Euclidean norm of $\ul x\in\CC^L$ is denoted by $\vecnorm{\ul x}$;  also, we let  $\abz{\ul x}$ be the vector whose $i$th element is  $|x_i|^2$.
We use $(\cdot)^T$, $(\cdot)^*$, and $(\cdot)^\dagger$ to indicate the transpose, complex conjugate, and conjugate transpose operators, respectively. Furthermore,
$\expe_{\mathbf{x}}\rlbo{\cdot}$  denotes expectation with respect to the random variable $\mathbf{x}$. The subscript will be omitted if obvious from the context. We use $\mathrm{Tr}[\cdot]$ to indicate the trace operator. 
We let $\jpg(\underline{0},\mathsf{K})$ be the multivariate complex-valued circularly  symmetric Gaussian distribution with covariance matrix $\mathsf{K}$. We use $\mathrm{unif}(a,b)$ to  denote a uniform distribution over the interval $[a,b)$.  
The indicator function is denoted by $\mathds{1}(\cdot)$. All  logarithms are in base two. 

\section{Channel Model}\label{s2}
Optical fiber systems employ optical amplification to compensate for losses in the fiber at the expense of an increased noise level.  Two amplification principles exist, namely, lumped and distributed amplification. 
Lumped amplification makes use of several amplifiers along the fiber. 
Distributed amplifications compensates for the energy loss continuously, so that the signal energy level remains roughly constant  throughout the propagation.  
Throughout the paper, we  consider the ideal distributed-amplification case, that is, the signal power is assumed to be constant throughout the propagation.

\subsection{Continuous-Time Model} For distributed amplification, the generalized NLSE captures the effect of nonlinearity, dispersion, and noise along the fiber \cite{Meccozzi_1994_jlt}. It is a nonlinear partial differential equation and can be written as
\begin{IEEEeqnarray}{c}
\frac{\partial  \mf a}{\partial z}+j\frac{\beta_2}{2}\frac{\partial ^2 \mf  a}{\partial t^2}-j\gamma | \mf a|^2  \mf a= \mf{n}.\label{manakov}
\end{IEEEeqnarray}
Here, $\gamma$ and $\beta_2$ are the nonlinear coefficient and the group-velocity dispersion parameter, respectively.  The variable $  \mathbf{ a}=\mathbf{ a}\lefto(z,t\right)$ indicates the complex envelope of the optical field in location $z$ and at time~$t$. 
Furthermore, $\mf n\lefto(z,t\right)$, which is a complex-valued zero-mean Gaussian process, models the amplification noise. 
This process is spatially white and its power spectral density $S_{\mf n}(f)$ is given by 
$
S_{\mf n}(f)=\tnp/\zf, 
$
where $\zf$ is the fiber length and $\tnp$ is the noise power spectral density at the receiver.
 Equation \eqref{manakov}, which unfortunately admits no analytical solution, can be regarded as a continuous-time channel model for a fiber-optical link, with input waveform $\mf a(0,t)$ and  output waveform $\mf a(Z,t)$.

\subsection{Discrete-Time Model} \label{dm}
We move from  continuous time to discrete time  by sampling the input signal every $\dt$ seconds. 
Through this sampling technique, we map  an input signal of duration $T-\dt$ seconds into a complex vector $\az{0}$ of dimension $L=T/\dt$.   
Similarly, at the receiver, we  sample the output signal and obtain a  complex vector.

The map between input and output vectors can be approximated by using the SSF method, which approximates the fiber-optical channel by a cascade of $K$  segments of  length $\dz=\zf/K$. 
For a fixed fiber length $Z$, the SSF method gets precise as $K$ goes to infinity (or, equivalently, as $\dz$ goes to zero). 
For $1\leq k\leq K$, we denote the output vector of the $k$th segment by $\az{k}=[\azt{k}{0}, \dots, \azt{k}{L-1}]^T$ (see Fig.~\ref{fig:SSFM}); also, we use $\az{0}$ to denote the input vector. 
The relations between the discrete-time  and the continuous-time channel inputs and outputs are
$	\azt{0}{l}=\mf a(0,l\dt)$ and
$	\azt{K}{l}=\mf a(\zf,l\dt),$
for $0\le l \le L-1$.

The output of each segment is computed by separating the linear and the nonlinear operations as illustrated in Fig.~\ref{fig:SSFM}.
Specifically, the evolution from $\az{k-1}$ to $\az{k}$ ($1\leq k\leq K$) involves the following three steps:

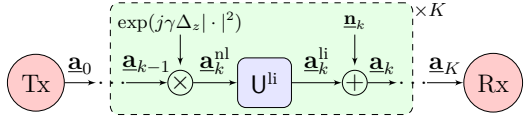
\begin{figure}[t]
	\begin{centering}
		\tikzstyle{block} = [rectangle, draw, fill=blue!10, 
		text width=1.5em, text centered, rounded corners, minimum height=2em]
		\tikzstyle{cloud} = [draw, circle,fill=red!20, node distance=3cm,
		minimum height=1em]
		\tikzstyle{crl} = [draw, circle, node distance=3cm,
		minimum height=1em]
		\tikzstyle{blockd} = [rectangle, dashed, draw, fill=green!10, 
		text width=11.5em, text centered, rounded corners, minimum height=4.5em]
		\tikzstyle{line} = [draw, -latex']
		\scalebox{0.8}{
			\begin{tikzpicture}[node distance = 3cm, auto]
			\node [cloud](Tr){Tx};
		\node[ right of=Tr, node distance=3.75cm](aad){};
			\node[blockd, above of=aad, node distance=.4cm](aa){};
		\node [crl,right of=Tr, node distance=2.4cm] (nl) {};
		\node (tms)[right of=Tr, node distance=2.4cm]{$\times$}(nl);
			\node [block, right of =nl, node distance=1.4cm] (li) {$\lo$};
			\node [crl, right of=li , node distance=1.5cm] (na) {};
					\node (tmse)[right of=li, node distance=1.5cm]{$+$}(na);
			\node [cloud, right of=na, node distance=2.3cm] (rec) {Rx};
					\node[ right of=Tr, node distance=6.55cm](aadd){};
		\node[ above of=aadd, node distance=1.2cm](adad){\footnotesize$\times K$};
			\path [line] (Tr) -- node{$\az{0}$}($(Tr)+(1,0)$);
			\node (bd)[right of =Tr, node distance=1.3cm]{$\dots$};
			\path[line]($(bd)+(0.2,0)$)--node{$\az{k-1}$}(nl);
			\path [line] (nl) -- node{$\an{k}$}(li);
			\path [line] (li) -- node{$\al{k}$}(na);
			\path[line](na)--node{$\az{k}$}(6,0);
			\path[line]($(tmse)+(1.2,0)$)--node{$\az{K}$}(rec);
			\path[line] ($(tms)+(0,.8)$)--(tms);
			\node [above of=tms,node distance=1cm]{\footnotesize$\mathrm{exp}(j\gamma\dz|\cdot|^2)$};
						\path[line] ($(tmse)+(0,.8)$)--(tmse);
	    	\node [above of=tmse,node distance=1cm]{\footnotesize$\nzn{k}$};
	    	\node [right of=tmse, node distance=1cm]{$\dots$};
			\end{tikzpicture}
		}
		\captionsetup{justification=centering}
		\caption{\small The channel model based on the split-step Fourier method.}\label{fig:SSFM}
	\end{centering}
\end{figure}

\begin{enumerate}
	\item \emph{Nonlinear step} (the Kerr effect): For $0\leq l \leq L-1$,
	\be\label{eq1625}
	\ant{k}{l}= \azt{k-1}{l}\, e^{j\gamma |\azt{k-1}{l}|^2\dz}.
	\ee
	\item  \emph{Linear step} (chromatic dispersion):
	\be
	\al{k}=\lo\an{k}.\label{eql}
	\ee
	Here, $\lo$ is a unitary matrix defined by
	\be\label{eql2}
	\lo=\ff^{\dagger} {\mathsf{D}^{\mathrm{li}}}\ff 
	\ee
	where $\ff$ is the discrete Fourier transform (DFT) operator with entries 
\begin{align}\label{dftdef}
f_{l,m}=\frac{1}{\sqrt{L}}e^{-j2\pi lm/L}\  , \ \ 0 \leq l,m \leq L-1.
\end{align}	
	Furthermore, ${\mathsf{D}^{\mathrm{li}}}=\text{diag}\rla{\dll{0}, \dots, \dll{L-1}}$ is a diagonal matrix with entries 
	\begin{align}\label{pll}
		\dll{l}=e^{j\dz\dllf\left(l\right)}, \quad l=0,\dots, L-1
	\end{align}
	where
	\begin{align}
		\dllf\lefto(l\right)&= 
		\frac{\beta_2}{2}\rlp{\frac{2\pi}{T}}^2\rlp{\frac{L}{2}-\left|\frac{L}{2}-l\right|}^2.\label{def:fli}
	\end{align}
	For efficient implementation, \eqref{eql2} is usually computed using the fast Fourier transform.
	\item \emph{Noise addition}: 
	\be
	\az{k}=\al{k} +\nzn{k}\label{an}
	\ee
	where $\nzn{k}\distas\jpg(\underline{0},\sigma_n^2 \mathsf{I}_L)$ 
	with 
	\begin{IEEEeqnarray}{rCL}\label{eq:sigma2}
		\sigma_n^2=\frac{\pn}{K}
	\end{IEEEeqnarray}
	where $\pn$ is the  per-sample noise variance, which can be calculated using the parameters of the  noise generated by the inline amplifiers as \cite{lau_2007_jlt}
	\begin{IEEEeqnarray}{rCl}
 \pn&=&\tnp B_n\\
 &=&h\nu Z \alpha n_{sp}B_n.\label{np}
	\end{IEEEeqnarray} 
	Here, $h\nu$ is the optical photon energy, $\alpha$ is the attenuation parameter,  $n_{sp}$ is the spontaneous emission factor, and $B_n$ is the receiver filter bandwidth. 
	Due to nonlinear effects, each signal frequency component interacts with all possible noise frequency components. However, this interaction becomes weaker as the frequency gap between these two components increases \cite{secondini2012analytical}.  
	  Here, we assume that the bandwidth of the receive filter is much greater than that of the signal (i.e., $B_n\dt\gg1$)  so that the influence of the interaction between the out-of-band noise and the signal can be neglected.
\end{enumerate}

Using $K$ times the three steps listed above, we obtain a probabilistic channel law that maps the input vector $\az{0}$ into the output vector $\az{K}$.
We shall refer to this law as the \emph{SSF channel} with length $\zf$ and number of segments $K$. 
Assuming that the SSF channel is block-memoryless across blocks of length $T$ seconds, we can write  its capacity (in bits per channel use) as 
\begin{IEEEeqnarray}{c}\label{eqc}
\cpct_K= \frac{1}{L}\sup I\lefto(\az{K};\az{0}\right).
\end{IEEEeqnarray}
 Here, the supremum is over all probability distributions on the input random vector $\az{0}$ that satisfy the power constraint $\expe\bigl[\vecnorm{\az{0}}^2\bigr]\leq L\po$, where $P$ is the input power.

The only known \emph{upper} bound on the capacity~\eqref{eqc} of the SSF channel is~\cite{Kramer_2015_itw,Yousefi_2015_cwit} 
\be\label{awgnub}
\cpct_K \leq \log\rlpo{1+\po/\pn}.
\ee
This bound is valid for every $K$. In contrast, a multitude of \emph{lower} bounds have been proposed. Most, if not all, such bounds use various mismatched decoding approaches, where nonlinear distortion is treated as noise at the receiver   \cite{mitra2001nonlinear,Ellis_2010_jlt,Meccozzi_2012_jlt,irukulapati2016tighter,secondini_2013_jlt}.

\section{  Lower Bounds on the Capacity of the SSF Channel}\label{s3}
In this section, we propose one simulation-based as well as two closed-form lower bounds on the capacity of the SSF channel, given in Theorems~1--3.
First, we lower-bound the capacity by a function that can be evaluated through  Monte Carlo simulation.
Second, we provide a lower bound on  this function by a closed-form expression to analyze our simulation results at high power. 
Third, we  replace our second bound by an explicit function of the input power and $K$ (the number of segments) at the expense of tightness.
We evaluate our first bound through simulations (see  Section~\ref{sec:simulation}), which indicate that this bound saturates in the high-power regime.  
Our second bound, although loose at low powers, can be used to approximate the asymptotes of the simulation results.  
Finally, we use our third bound to show that these asymptotes go to infinity as $K$ grows large.
We use the following two lemmas to establish our first lower bound in Theorem~\ref{thm1}.
\begin{lemma}\label{lem:iidg}\normalfont
	If the input vector distribution of the SSF channel is i.i.d. Gaussian, i.e., $\az{0}\distas~\jpg(\underline{0},\po \mathsf{I}_L) $,  then
	\be\az{K}\distas\jpg(\underline{0},(\po+\pn) \mathsf{I}_L).\label{eq:l1}\ee
\end{lemma}
\begin{IEEEproof}
	See Appendix~\ref{aps}.
\end{IEEEproof}
\begin{lemma}\label{lema2}\normalfont
	Let $(\xx,\yy)$ be a pair of $L$-dimensional proper complex random vectors, distributed according to an arbitrary joint probability density function. The conditional entropy $h(\yy\given\xx)$ is bounded as
	\be\label{eql1}
	h\lefto(\yy\big|\xx\right)\leq \frac{L}{2}\log\lefto(2\pi^3 e\frac{\kappa(\yy\given \xx)}{L}\right)
	\ee
where
	\begin{IEEEeqnarray}{rCL}\label{eq:kappa1}
		\kappa(\yy\given \xx)=\sum\limits_{i=0}^{L-1}\expe\rlbo{\absl{{\mf y_i}}^4}-\sum\limits_{i=0}^{L-1}\expe_{\xx}\rlbo{\expe_{\yy}\rlbo{\absl{\mf y_i}^2| \xx}^2}.\IEEEeqnarraynumspace
	\end{IEEEeqnarray}

\end{lemma}
\begin{IEEEproof}
	See Appendix~\ref{apB}.
\end{IEEEproof}
Based on these two lemmas, we can now formulate our first lower bound as follows.
\begin{theorem}\label{thm1}\normalfont
	
	Let $\az{0}\distas \jpg\rlpo{\ul{0},\po\mathrm{I}_L}$. 
	The capacity \eqref{eqc} of the SSF channel  is lower-bounded as
	\begin{IEEEeqnarray}{c}
	\cpct_K\geq	\lone
		= \frac{1}{2}\log\rlp{\frac{e}{2\pi}\cdot\frac{\rlp{\po+\pn}^2}{2\rlp{\po+\pn}^2-\mathcal{E}}}\label{kappabound}
	\end{IEEEeqnarray}
\end{theorem}
where 
\be\label{E}
\mathcal{E}=\frac{1}{L}\sum\limits_{i=0}^{L-1}\expe_{\az{0}}\rlbo{\expe\rlbo{|\azt{K}{i}|^2|\az{0}}^2}.
\ee
\begin{IEEEproof}
	By  \eqref{eqc}, we have that $\cpct_K\geq I\rlpo{\az{K};\az{0}}/L$, where $\az{0}\distas \jpg\rlpo{\ul{0},\po\mathrm{I}_L}$. 
Next, we  decompose $I(\az{K};\az{0})$ as  
	\be\label{I}
	I\rlpo{\az{K};\az{0}}=h(\az{K})-h(\az{K}|\az{0}).
	\ee
	For our choice of input distribution, Lemma~\ref{lem:iidg} yields that  $\az{K}\distas \jpg\rlpo{\ul{0}, \rlp{\po+\pn}\mathrm{I}_L}$. Hence, we conclude that \cite[Thm.~2]{Neeser_1993_tinf} 
	\begin{IEEEeqnarray}{c}
		h(\az{K})=L\log\rlpo{\pi e\rlp{\po+\pn}}\label{h}.
	\end{IEEEeqnarray}
	Furthermore, it follows from Lemma~\ref{lema2} that
	\be\label{eq22}
	h\rlpo{\az{K}|\az{0}}\leq \frac{L}{2}\log\rlpo{2\pi^3 e \frac{\kappa\rlpo{\az{K}|\az{0}}}{L}},
	\ee
	where, by definition (see \eqref{eq:kappa1}),
		\begin{IEEEeqnarray}{rCL}\label{eq:kappa2}
			\kappa(\az{K}\given \az{0})&=&\sum\limits_{i=0}^{L-1}\expe\rlbo{\absl{\azt{K}{i}}^4}-\sum\limits_{i=0}^{L-1}\expe_{\az{0}}\rlbo{\expe_{\az{K}}\rlbo{\absl{\azt{K}{i}}^2| \az{0}}^2}\\
			&=& \rlp{2(P+\pn)^2- \mathcal{E}}L.\label{eq:kapa:e}
		\end{IEEEeqnarray}
		Here, in the last step we used that  $\azt{K}{i}\distas \mathcal{CN}\rlpo{0,P+\pn}$.
	Substituting \eqref{eq:kapa:e} into \eqref{eq22} and then \eqref{h} and \eqref{eq22} into \eqref{I}, we obtain \eqref{kappabound}.
\end{IEEEproof}
In the absence of a closed-form expression, \eqref{kappabound} can be calculated by evaluating  $\mathcal{E}$
through Monte Carlo simulation (see Section~\ref{sec:simulation}). 
Further lower-bounding $\lone$ in \eqref{kappabound}, one can obtain an expression that can be evaluated in closed form, as shown in the next theorem.

\begin{theorem}\label{tm2}\normalfont
For every $0\leq m\leq L-1,$ 
let
	\begin{IEEEeqnarray}{rCl}
	\alf{0}{m}&=&\frac{1}{\nt}\sum\limits_{l=0}^{\nt-1}e^{j\dz\dllf\left(l\right)}e^{-j2\pi lm/L}\label{Alfdef}\\
	\alpha&=&\rlp{\sum\limits_{\substack{m=0}}^{L-1}|\alf{0}{m}|}^2-\sum\limits_{\substack{m=0}}^{L-1}|\alf{0}{m}|^2\label{alfdef}\\
	\UB&=&\alpha K\rlp{ \frac{K^3}{2e\gamma^2 \pn Z^2}+\frac{K}{eZ\gamma}+\frac{\pn}{2eK}}\label{Mkdef}
		\end{IEEEeqnarray}
		and
			\begin{IEEEeqnarray}{rCl}
	\zeta(\po,K)&=&|\alf{0}{0}|^{4(K-1)}\mathrm{exp}\rlpo{-\frac{\UB}{P|\alf{0}{0}|^{2(K-1)}}}\label{zeta}\IEEEeqnarraynumspace
	\end{IEEEeqnarray}
	Then $\lone$ in~\eqref{kappabound}  can be  lower-bounded as  
	\begin{IEEEeqnarray}{c}\label{alphabound}
	\lone \geq \ltwo= \frac{1}{2}\log\rlp{\frac{e}{4\pi}\cdot\frac{(\po+\pn)^2}{(\po+\pn)^2-\zeta\rlpo{\po,K}\po^2}}.\IEEEeqnarraynumspace
	\end{IEEEeqnarray}

\end{theorem}\normalfont
\begin{IEEEproof}
By comparing \eqref{kappabound} and \eqref{alphabound}, we see that, to prove Theorem~\ref{tm2}, it is sufficient to show that $\mathcal{E}\ge 2P^2\zeta\rlpo{P,K}$, where $\mathcal{E}$ is defined in \eqref{E}.  In Appendix~\ref{apC}, we prove that for every $0\leq r\leq L-1$ 
	\begin{align}
		\ea\rlbo{\rlp{\expe_{\az{K}}\rlbo{|\azt{K}{r}|^2\given\az{0}}}^2}&\geq 2P^2\zeta\rlpo{\po,K}\label{eqappb}
	\end{align}
	from which the desired result follows. 
\end{IEEEproof}
For any given $K$, \eqref{alphabound} yields
\begin{align}
	\lim\limits_{\po\to \infty} \ltwo = -\frac{1}{2}\log\rlp{\frac{4\pi}{e}\rlp{1-|\alf{0}{0}|^{4(K-1)}}}=	\ltwoinf.\label{asympa}
\end{align}
Note that $|\alf{0}{0}|\leq 1$ which follows by triangle inequality, so the expression on the RHS of \eqref{asympa} is well defined. The limit in \eqref{asympa} reveals that $\ltwo$ approaches the constant $\ltwoinf$ as $P\to \infty$,  and the value of this constant  depends on $K$. We are interested in how $\ltwo$ and its asymptote limit $\ltwoinf$ behave as $K$ grows large. This behavior will be illustrated numerically in Section~\ref{sec:simulation}, where we use  $\ltwoinf$ to provide a lower bound on the high-power asymptote of $\lone$. In Theorem~\ref{th3} we further lower-bound $\ltwo$ to obtain a less tight but simpler bound on $\mathcal{C}_K$, which reveals its dependence on $K$ in the asymptotic limit $\po\to\infty$.
\begin{theorem}\label{th3}\normalfont
	Let 
	\begin{IEEEeqnarray}{rCl}
		\Cone&=&\frac{Z^2}{L}\sum\limits_{l=0}^{L-1}{\rlp{\dllf\rlpo{l}}^2},\label{c1def}\\
		\Ctwo &=&\sqrt{6}{Z}\sum\limits_{l=0}^{\nt-1}\dllf\left(l\right)\label{c2def},
		\end{IEEEeqnarray}
		and
			\begin{IEEEeqnarray}{rCl}\label{semosalas}
		\UBtw&=&\Ctwo\rlp{\frac{\Ctwo }{4K}+1}\rlp{ \frac{K^3}{2e\gamma^2 \pn Z^2}+\frac{K}{eZ\gamma}+\frac{\pn}{2eK}}.
	\end{IEEEeqnarray}
	Then, for every  $K>\max\rla{\frac{\beta_2 Z\pi^2}{2\sqrt{2}\dt^2},\sqrt{\Cone}}$, the bound $\ltwo$ in \eqref{alphabound} is further lower-bounded as
	\begin{IEEEeqnarray}{c}
	\ltwo\geq\lthree=\frac{1}{2} \log\rlpo{\frac{e}{4\pi}\cdot\frac{ (\po+\pn)^2}{2\Cone \po^2/K+2 \po\pn+\po \UBtw+\pn^2}}.\label{cbound}\IEEEeqnarraynumspace
	\end{IEEEeqnarray}

\end{theorem}
\begin{IEEEproof}
	See Appendix~\ref{apD}.
\end{IEEEproof}
We use $\lthree$ to lower-bound the asymptotic behavior of $\lone$ and $\ltwo$ as $\po\to\infty$. Since we are interested in these asymptotes as $K$ grows large,   the condition on $K$ mentioned in Theorem~\ref{th3} is not restrictive.
We obtain from \eqref{cbound} that
\begin{align}\label{asympc}
	\lim\limits_{\po\to \infty} \lthree = \frac{1}{2}\log\rlpo{\frac{eK}{8\pi\Cone}}=\lthreeinf.
\end{align} 
To be more specific, it follows from \eqref{cbound} that if $K$ grows with $P$ so that $P/K^3\to\infty$, then the lower bound $\lthree$ goes to infinity.
As  mentioned in Section~\ref{sec:intro}, the SSF method with a fixed number of segments is a valid approximation of NLSE up to a certain power.
Our lower bound $\lthree$ indicates that if we increase the number of segments $K$ with power $P$ such that $P/K^3\to\infty$ as $P\to\infty$, the capacity grows unboundedly.

\section{Numerical Example}\label{sec:simulation}

\begin{figure}[!t]
	\centering
\includegraphics{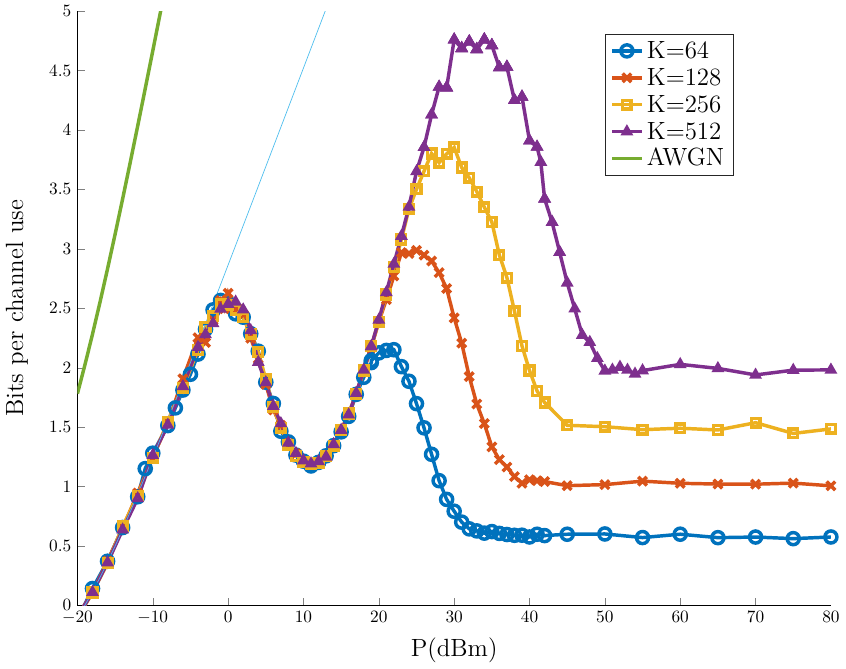}
\caption{\small Simulation results for  the lower bound $\lone$ in \eqref{kappabound}, for a different numbers of channel segments $K$. The AWGN upper bound in \eqref{awgnub} and the low-power approximation $L^{(1)}_{\mathrm{LP}}$ in \eqref{kappabound_simplef} are also shown.}
\label{fig_sim}
\end{figure}
\begin{table}[!t]
	\renewcommand{\arraystretch}{1.3}
	\caption{Channel parameters used in the example.}
	\label{t2}
	\centering
	\begin{tabular}{c c c }
		\hline
		\hline
		Parameter&Symbol& Value\\
		\hline
		Fiber length & $Z$&$850 \ \mathrm{km}$\\
		Attenuation &$\alpha$ & $0.2 \ \mathrm{dB/km}$\\
		Dispersion& $\beta_2$ & $-21.7 \ \mathrm{ps^2/km}$\\
		Nonlinearity& $\gamma$ & $1.27 \ \rlp{\mathrm{W km}}^{-1}$\\
		Symbol time & $\dt$&$100 \ \mathrm{ps}$\\
		Optical photon energy & $h\nu$&$1.3\cdot 10^{-19} \ \mathrm{J}$\\
		Spontaneous emission factor & $n_{sp}$&$4$\\
		Filter bandwidth  & $B_n$&$200  \ \mathrm{GHz}$\\
		\hline
		\hline
	\end{tabular}
\end{table}

In this section, we present and analyze the results obtained by evaluating $\lone$ in \eqref{kappabound} through Monte Carlo simulation. 
After stating the channel parameters, we analyze our simulation results in low, moderate, and high power regimes separately. Finally, we draw  conclusions based on our analytical and numerical results.

We consider  a single-mode fiber link  with parameters given in Table~\ref{t2}.  The per-sample noise variance  can be calculated through \eqref{np} and is equal to  $\pn=4.1 \ \mu\text{W}$.
Four different values of channel segments $K$ are considered. They correspond to segment lengths $\dz$ of (approximately)  $13.3$, $6.6$, $3.3$, and $1.7$ km.

In Fig.~\ref{fig_sim}, $\lone$ is numerically evaluated by Monte Carlo simulation. For every  $K$ and $P$, 200 independent realizations of $\az{0}$,  with length $L=2000$, were generated, and for each of these realizations, 1000 independent realizations of $\az{K}$ were generated using \eqref{eq1625}--\eqref{an}. 

As can be seen in Fig.~\ref{fig_sim}, in the low-power regime the evaluation of $\lone$ results in the same lower bound for all the four considered values of $K$. This is because at  low power, the SSF method  models the NLSE  accurately for all values of $K$ considered here. Since in low-power regime the nonlinearity can be neglected, it is possible to obtain a closed-form accurate approximation of $\lone$ in this regime. Specifically, by setting $\gamma=0$, the SSF channel turns into the  linear channel $\az{K}=(\lo)^K\az{0}+\mathbf{n}$, where $\mathbf{n}\distas \mathcal{CN}\rlpo{0,\pn\mathsf{I}_L}$. Noting that $(\lo)^K\az{0}\distas \mathcal{CN}(0,P\mathsf{I}_L)$,  one can evaluate $\mathcal{E}$ for this channel in closed form and  obtain
\begin{IEEEeqnarray}{c}
	L^{(1)}_{\mathrm{LP}}=\frac{1}{2}\log\rlp{\frac{e}{2\pi}\rlp{1+\frac{\po^2}{\rlp{2\po+\pn}\pn}}}.\label{kappabound_simplef}
\end{IEEEeqnarray}
  This approximation, which is plotted in Fig.~\ref{fig_sim},  is accurate for values of power $P$ less than 0 dBm.  
 
At moderate power levels, our bound shows a peak at approximately  0 dBm.  We next provide an intuitive discussion to explain why our bound decreases in the interval $[0\  \text{dBm}\ , \ 10  \ \text{dBm}]$. At moderate power levels,  the effects of the nonlinearity become substantial. The interaction between the nonlinearity and the noise  changes the phase of the signal randomly during propagation.
This phase noise  leads to  amplitude noise when the chromatic dispersion is applied. Next, we show by an example that having amplitude randomness at the receiver causes an increase of  $\kappa(\az{K}|\az{0})$ and hence a decrease of $\lone$. Define a random vector $\azexp{K}$ satisfying  $|\azexpt{K}{i}|^2=|\azt{K}{i}|^2+\npt{i}$ for $i=0,\dots L-1$, where $\np$ is a signal-independent zero-mean noise with covariance matrix $\pnexp\mathsf{I}_L$. Using the definition of $\kappa(\cdot|\cdot)$ in \eqref{eq:kappa1}, we have
\begin{IEEEeqnarray}{rCl}
	\kappa\rlpo{\azexp{K}|\az{0}}=\kappa\rlpo{\az{K}|\az{0}}+L\pnexp.
\end{IEEEeqnarray}
As shown in Fig.~\ref{fig_sim}, our bound starts increasing again roughly at 10 dBm. 

\begin{figure}[!t]
	\centering
	\includegraphics{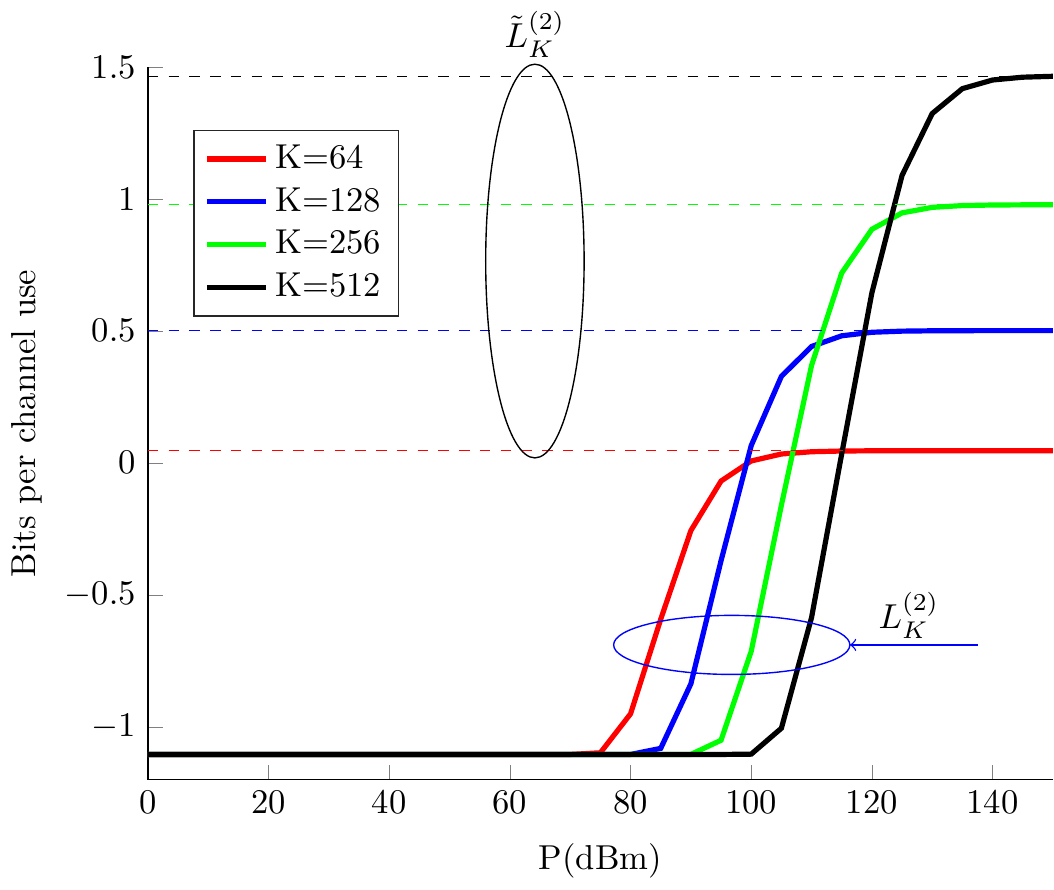}
	\caption{\small Numerical evaluation of the lower bound $\ltwo$ in \eqref{alphabound}, for a different number of channel segments $K$. The high-power asymptote $\ltwoinf$ in \eqref{asympa} is also illustrated by horizontal dashed lines.}
	\label{fig_sim2}
\end{figure}

In the high-power regime, as  can be seen in Fig.~\ref{fig_sim},  $\lone$ becomes sensitive to $K$.
This is due to the fact that for a fixed $K$, the SSF method is accurate up to a certain power. In other words, as far as the calculation of $\lone$ is concerned, the SSF method with $K=64$, $128$, and $256$ segments accurately models the continuous channel only up to the power levels of $19$, $22$, and $25$ dBm, respectively.  
    
As it is evident in Fig.~\ref{fig_sim}, $\lone$ eventually saturates at high power levels, and the saturation point increases with $K$. We use our asymptotic closed-form bound $\ltwoinf$ in \eqref{asympa} to approximate these asymptotes.
 In Fig.~\ref{fig_sim2}, the lower bound $\ltwo$  is evaluated as a function of power for different numbers of channel segments $K$; furthermore, the asymptote of this lower bound, $\ltwoinf$, is also shown via horizontal dashed lines.
Comparing the results in Fig.~\ref{fig_sim} and Fig.~\ref{fig_sim2}, one observes that 
  approximations made by $\ltwoinf$ (which are $0.05$, $0.5$, $0.98$, and $1.47$ ) are different from the simulation asymptotes (which are $0.58$, $1$, $1.49$, and $1.98$) by a constant value of approximately $0.5$ bits, which becomes negligible as  $K$ becomes large. The asymptotic lower bound $\lthreeinf$ \eqref{asympc}  suggests that the saturation point increases by $0.5$ bits if the value of $K$ doubles, which is evident in Figs. \ref{fig_sim} and \ref{fig_sim2}.

  To summarize, we evaluated $\lone$ through simulation and observed that this lower bound increases with $K$. Next, we used our analytical bound $\lthreeinf$ to show that this trend will sustain for large values of $K$. Therefore, one may conclude that, as long as the effects of spectrum broadening are neglected, the capacity of the NLSE channel goes to infinity with power.

\section{Conclusion}

We presented a lower bound on the capacity of the split-step Fourier method channel, which can be evaluated by calculating a double expectation using Monte Carlo simulations. Doing so, we evaluated this bound for different  numbers of channel segments $K$, and different transmit power levels. Simulation results indicated that for a fixed $K$, the lower bound saturates at high power and the saturation point increases with $K$. To study the asymptotic behavior of this bound, we further lower-bounded it by  two closed-form expressions. Our analytical results prove that with appropriate choices of power $P$ and number of segments $K$, the capacity of the SSF channel can be made arbitrarily large.  Using our  analytical bounds, we proved that the saturation point  increases to infinity as we increase the number of channel segments. Specifically, we showed that the asymptotes of our bound increase by 0.5 bit if the number of segments is doubled. Our numerical and analytical results suggest that as long as the effect of spectrum broadening is ignored, the capacity of the fiber-optical channel described by the NLSE  goes to infinity with power.

\appendices
\def\thesubsectiondis{\thesection.\Roman{subsection}.} 
\renewcommand\thesubsection{\thesection.\Roman{subsection}}
	\section{Preliminaries}
	\subsection{Maximum Entropy}
	Among all real random vectors $\xx$ with a fixed nonsingular  correlation matrix ${\R\lefto(\xx\right)=\expe\lefto[\xx \xx^T\right]}$, the joint Gaussian distribution has  maximum differential entropy \cite[Thm. 8.6.5]{cover_information2}, i.e., 
	\begin{IEEEeqnarray}{c}\label{w}
	h\lefto(\xx\right)\leq \frac{1}{2}\log\rlpo{\left(2\pi e\right)^L \det \R\lefto(\xx\right)}.
	\end{IEEEeqnarray}
	 Using Hadamard's inequality \cite[Thm.~17.9.2]{cover_information2} and  Jensen's inequality \cite[Sec. 2.6]{cover_information2} we can further upper-bound \eqref{w} as~\cite{Kramer_2015_itw}
	\begin{IEEEeqnarray}{rCl}
	h\lefto(\xx\right)&\leq& \frac{L}{2}\log\rlpo{\frac{2\pi e}{L} \Tr\bigl[\mathsf{R}\lefto(\xx\right)\bigr]}\label{eq21102}\\
	&=&\frac{L}{2}\log\rlpo{\frac{2\pi e}{L} \expe\bigl[\vecnorm{\xx}^2\bigr]}.\label{twostar}
	\end{IEEEeqnarray}
	
	\subsection{Polar Coordinate System}
	The differential entropy of a complex random variable $\mf x$ can be computed in polar coordinates as \cite[Lemma 6.16]{lapidoth_2003_tinf}
	\be
	h\lefto(\mathbf x\right)=h\lefto(\absol{\mathbf x},\ang{\mathbf{x}}\right)+\expe\rlbo{\log\absol{\mathbf x}}.\label{eq1940}
	\ee
	Here, $\ang{\mf x}$ denotes the phase of $\mf x$.
	Furthermore, \cite[Lemma 6.15]{lapidoth_2003_tinf}
	\begin{IEEEeqnarray}{c}\label{d}
	h\rlpo{|\mf x|^2}=h(|\mf x|)+\expe\rlbo{\log|\mf x|}+\log 2.
	\end{IEEEeqnarray}
	Using \eqref{eq1940} and \eqref{d}, we can upper-bound $h(\mf x)$ as 
	\begin{IEEEeqnarray}{rCl}
		h\lefto(\mathbf{x}\right)&=&h\rlpo{|\mathbf{x}|}+h\rlpo{\ang{\mathbf{x}}\left||\mathbf{x}|\right.}+\expe\rlbo{\log|\mathbf{x}|}\\
		&=&h\rlpo{|\mathbf{x}|^2}-\log 2+h\rlpo{\ang{\mathbf{x}}\left||\mathbf{x}|\right.}\\
		&\leq& h\lefto(\absol{\mathbf{x}}^2\right)+\log\pi.
	\end{IEEEeqnarray}
	In the last step, we used that $h(\ang{\mf x}\given |\mf x|)\leq h(\ang{\mf x})\leq \log 2\pi$.
	Extending this inequality to $L$-dimensional complex random vectors, we obtain
	\be\label{eq2050}
	h\lefto(\xx\right)\leq L\log\pi+h\lefto(\abx{\ul{\mathbf{x}}}\right).
	\ee
\section{Proof of Lemma~\ref{lem:iidg}}\label{aps}
Since the nonlinear step is memoryless (see \eqref{eq1625}), $\an{1}$ remains i.i.d.. Also, since the phase $\ang{\azt{0}{r}}$ of $\azt{0}{r}$, $\ r=0,\dots,L-1$, is distributed uniformly over the interval $[-\pi , \pi)$  and is independent of $|\azt{0}{r}|$, the random variable $\ang{\ant{1}{r}}$, which is equal to $\ang{\azt{0}{r}}+\dz \gamma |\azt{0}{r}|^2$, 
 is also uniformly distributed over $[-\pi, \pi)$ and  independent of $|\ant{1}{r}|$. Since $|\ant{1}{r}|=|\azt{0}{r}|$, we  conclude that  $\an{1}\distas \jpg(\underline{0},\po \mathsf{I}_L)$. Furthermore, since $\lo$ is a unitary matrix,
$\al{1}\distas \jpg\rlp{\underline{0},\po \mathsf{ I}_L}.$
Therefore, after noise addition, we conclude that $\az{1}\distas  \jpg\rlp{\underline{0},(\po+\sn^2) \mathsf{ I}_L}.$
Repeating the same calculations $K$ times and using \eqref{eq:sigma2}, we obtain \eqref{eq:l1}.

\section{ Proof of Lemma \ref{lema2}}\label{apB}
It follows by a conditional version of~\eqref{eq2050} that
\begin{IEEEeqnarray}{rCL}
	\IEEEeqnarraymulticol{3}{l}{h\lefto(\yy|\xx=\ul x\right)}\nonumber\\
	&\leq & L\log\pi+h\lefto(\left.\abz{ \yy}\middle|\xx=\ul x\right.\right)\\
	& = & L\log\pi+h\lefto(\left.  \abz{\yy}-\expe_{\yy}\rlbo{\left.\abz{  \yy}\right|\xx=\ul x}\right|\xx=\ul x\right).\IEEEeqnarraynumspace  
\end{IEEEeqnarray}
The last step follows because differential entropy is invariant to translations \cite[Thm. 8.6.3]{cover_information2}.
For every $\ul x\in\CC^L$, we define the random vector $\ww(\ul x)=\abz{ \yy}-\expe\rlbo{\left. \abz{ \yy}\right|\xx=\ul x}$. Since $\ww(\ul x)$ has real entries, we can use~\eqref{twostar} to obtain
\begin{IEEEeqnarray}{c}\label{21}
	h\lefto(\yy|\xx=\ul x\right) \leq \frac{L}{2}\log\rlpo{\frac{2\pi^3 e}{L}\expe_{\yy}\rlbo{\vecnorm{\ww(\ul x)}^2\big|\xx=\ul x}}.\IEEEeqnarraynumspace
\end{IEEEeqnarray}
Averaging both sides of~\eqref{21} with respect to $\xx$, we obtain
\begin{IEEEeqnarray}{rCL}
	h\lefto(\yy\big|\xx\right)&=&\frac{L}{2}\expe_{\xx}\rlbo{\log\rlpo{\frac{2\pi^3 e}{L}\expe_{\yy}\rlbo{\vecnorm{\ww(\xx)}^2}}}\IEEEeqnarraynumspace\\
	&\leq& \frac{L}{2}\log\rlpo{\frac{2\pi^3 e}{L}{\expe_{\xx}\rlb{\expe_{\yy}{\rlb{\vecnorm{\ww(\xx)}^2}}}}}.
\end{IEEEeqnarray}
Here, the last inequality follows from Jensen's inequality. To conclude the proof, we note that
\begin{IEEEeqnarray}{rCL}
	\expe\rlbo{\vecnorm{\ww(\xx)}^2}&=&\expe\rlbo{\sum\limits_{i=0}^{L-1}\rlp{\left.|\mf y_i|^2-\expe\rlbo{|\mf y_i|^2\right| \xx}}^2}\\
	&=&\sum\limits_{i=0}^{L-1}\expe\rlbo{|\mf y_i|^4}-\expe_{\xx}\rlbo{{\expe_{\yy}\rlbo{\left.|\mf y_i|^2\right| \xx}}^2}\IEEEeqnarraynumspace\\
	&=&\kappa\rlpo{\left.\yy\right| \xx}.
\end{IEEEeqnarray}

	\section{Proof of \eqref{eqappb}}\label{apC}

	To prove \eqref{eqappb}, we start by noting that (see Fig.~\ref{fig:SSFM})
	\begin{IEEEeqnarray}{rCl}
	\expe\rlbo{\left.|\azt{K}{r}|^2\right|\az{0}}&=&\expe\rlbo{\left.|\alt{K}{r}|^2\right|\az{0}}+\sigma^2_n\label{p0}\\
	&\geq& \expe\rlbo{\left.|\alt{K}{r}|^2\right|\az{0}}.\label{eqnf1}
	\end{IEEEeqnarray}
	
	In Appendix \ref{sec1}, we prove that
	for every $2\leq k\leq K,$
	\begin{IEEEeqnarray}{c}
	\expe\rlb{\left.\absol{\alt{k}{r}}^2\right|\az{0}}\geq \expe\rlbo{\left.\absol{\azt{k-1}{r}}^2\right|\az{0}}\absol{\alf{0}{0}}^2-\frac{\UB}{K}\IEEEeqnarraynumspace\label{77}
\end{IEEEeqnarray}
where $\alf{0}{0}$ and $\UB$   are defined in  \eqref{Alfdef} and  \eqref{Mkdef}, respectively.
Using  \eqref{eqnf1} and \eqref{77} $K-1$ times, we obtain (recall that $\alt{1}{r}$ is a deterministic function of $\az{0}$)
\begin{IEEEeqnarray}{c}
\expe\rlbo{\left.|\azt{K}{r}|^2\right|\az{0}}\geq|\alt{1}{r}|^2\absol{\alf{0}{0}}^{2(K-1)}-\frac{\UB}{K}\sum\limits_{j=0}^{K-2}\absol{\alf{0}{0}}^{2j}.\label{yes}
\end{IEEEeqnarray}
Furthermore, since $|\alf{0}{0}|\leq1$ (see \eqref{Alfdef}), we conclude that
\begin{IEEEeqnarray}{c}
\expe\rlbo{\left.|\azt{K}{r}|^2\right|\az{0}}
\geq|\alt{1}{r}|^2\absol{\alf{0}{0}}^{2(K-1)}-\UB.\label{dddesham}
\end{IEEEeqnarray}
Squaring both sides of \eqref{dddesham},   
we get
\begin{IEEEeqnarray}{c}
\rlp{\expe\rlbo{\left.|\azt{K}{r}|^2\right|\az{0}}}^2
\geq\rlp{|\alt{1}{r}|^2\absol{\alf{0}{0}}^{2(K-1)}-\UB}^2 \times \mathds{1}\rlpo{|\alt{1}{r}|^2>{\UB}/{\absol{\alf{0}{0}}^{2(K-1)}}}\label{qeq83}.\IEEEeqnarraynumspace
\end{IEEEeqnarray}
We know from the proof of Lemma~\ref{lem:iidg} that $ \al{1}\distas\jpg(\underline{0},P \mathsf{I}_L)$. Consequently,  ${|\alt{1}{r}|^2\distas \text{Exp}\rlpo{1/\po}}$. Averaging both sides of \eqref{qeq83} with respect to $\az{0}$, we obtain 
\begin{IEEEeqnarray}{rCL}
\expe\rlb{\rlp{\expe\rlbo{\left.|\azt{K}{r}|^2\right|\az{0}}}^2}&\geq&\frac{1}{\po}\int\limits_{{\UB}/{\absol{\alf{0}{0}}^{2(K-1)}}}^{\infty}\rlp{\ai\absol{\alf{0}{0}}^{2(K-1)}-\UB}^2\exp\rlpo{-\frac{\ai}{\po}}\di\ai\label{q83}\\
&=& 2\po^2\absol{\alf{0}{0}}^{4(K-1)}\exp\rlpo{\frac{-\UB}{\po\absol{\alf{0}{0}}^{2(K-1)}}}\\
&=& 2\po^2\zeta\rlpo{P,K}\label{eq84}
\end{IEEEeqnarray}
where in the last step we used \eqref{zeta}.

\subsection{Proof of \eqref{77} }\label{sec1}

Let $\anf{k}=\ff\an{k} $. By definition of the DFT operator $\ff$ in \eqref{dftdef}, we have
\be\antf{k}{n}=\frac{1}{\sqrt{\nt}}\sum\limits_{m=0}^{\nt-1} \ant{k}{m}e^{-j\frac{2\pi}{\nt}mn}.\label{eqn52}\ee
Furthermore, it follows from \eqref{dftdef} and \eqref{pll} that 
\begin{align}
	\alt{k}{r}&=\frac{1}{\sqrt{\nt}}\sum\limits_{n=0}^{\nt-1}\antf{k}{n} e^{j\frac{2\pi}{\nt}nr}e^{j\dz\dllf(l)}\\
	&=\frac{1}{{\nt}}\sum\limits_{n=0}^{\nt-1}\sum\limits_{m=0}^{\nt-1}\ant{k}{m}e^{j\frac{2\pi}{\nt}n\left(r-m\right)}e^{j\dz\dllf(l)}\\
	&=\frac{1}{{\nt}}\sum\limits_{m=0}^{\nt-1}\ant{k}{m}\sum\limits_{n=0}^{\nt-1}e^{j\frac{2\pi}{\nt}n\left(r-m\right)}e^{j\dz\dllf(l)}\\
	&=\sum\limits_{m=0}^{\nt-1}\ant{k}{m}\alf{r}{m}\label{fdddd}
\end{align}
where 
\begin{IEEEeqnarray}{rCl}\label{eq:alf}
	\alf{r}{m}=\frac{1}{L}\sum\limits_{n=0}^{L-1}e^{j\frac{2\pi}{L}n(r-m)}e^{j\dz\dllf(l)}.
\end{IEEEeqnarray}
By squaring both sides of \eqref{fdddd}, we obtain
\begin{IEEEeqnarray}{c}\label{eqn541}
\absol{\alt{k}{r}}^2=\sum\limits_{m=0}^{L-1}|\ant{k}{m}|^2|\alf{r}{m}|^2+\sum\limits_{\substack{m,n\\m\neq n}}\ant{k}{m}\ants{k}{n}\alf{r}{m}\alf{r}{n}^*.
\end{IEEEeqnarray}
Taking the expectation of both sides of \eqref{eqn541} and using  triangle inequality, we get
\begin{IEEEeqnarray}{rCl}
\expe\rlb{\left.\absol{\alt{k}{r}}^2\right|\az{0}}&=&\sum\limits_{m}\expe\rlbo{|\ant{k}{m}|^2}|\alf{r}{m}|^2+\sum\limits_{\substack{m,n\\m\neq n}}\expe\rlbo{\ant{k}{m}\ants{k}{n}}\alf{r}{m}\alf{r}{n}^*\\
&\geq&\expe\rlbo{\left.\absol{\ant{k}{r}}^2\right|\az{0}}\absol{\alf{r}{r}}^2-\big|\sum\limits_{\substack{m,n\\m\neq n}}\expe\rlbo{\ant{k}{m}\ants{k}{n}}\alf{r}{m}\alf{r}{n}^*\big|\\
&\geq& \expe\rlbo{\left.\absol{\ant{k}{r}}^2\right|\az{0}}\absol{\alf{r}{r}}^2
-\sum\limits_{\substack{m,n\\m\neq n}}\left|\expe\rlbo{\left.\ant{k}{m}\ants{k}{n}\right|\az{0}}\right||\alf{r}{m}||\alf{r}{n}|\label{q02}\\
&\geq& \expe\rlbo{\left.\absol{\ant{k}{r}}^2\right|\az{0}}\absol{\alf{r}{r}}^2-\max_{m,n}\rla{\left|\expe\rlbo{\left.\ant{k}{m}\ants{k}{n}\right|\az{0}}\right|}\sum\limits_{\substack{m,n\\m\neq n}}|\alf{r}{m}||\alf{r}{n}|.\label{sar}
\end{IEEEeqnarray}
We note that since $\alf{r}{n}$ depends on $(r,m)$ only through $r-m$ (see \eqref{eq:alf}), for every $r=0, \dots, L-1$, we have 
\be\label{alfrrr}
\absol{\alf{r}{r}}^2=\absol{\alf{0}{0}}^2
\ee
 and
\begin{IEEEeqnarray}{rCl}\label{eq:alf:sum}
\sum\limits_{\substack{m,n\\m\neq n}}|\alf{r}{n}||\alf{r}{m}|&=&\sum\limits_{\substack{m,n\\m\neq n}}|\alf{0}{n}||\alf{0}{m}|\\
&=&\rlp{\sum\limits_{\substack{m=0}}^{L-1}|\alf{0}{m}|}^2-\sum\limits_{\substack{m=0}}^{L-1}|\alf{0}{m}|^2=\alpha
\end{IEEEeqnarray}
where in the last step we used \eqref{alfdef}.
In Appendix \ref{sss}, we prove that for all $2\leq k\leq K$ and all $0\leq m,n\leq L-1$,  
\be\label{asin}
\left|\expe\rlbo{\left.\ant{k}{m}\ants{k}{n}\right|\az{0}}\right|\leq   \frac{K^3}{2e\gamma^2 \pn Z^2}+\frac{K}{eZ\gamma}+\frac{\pn}{2eK}=\frac{\UB}{\alpha K}
\ee
where in the last step we used \eqref{Mkdef}.
Since $|\ant{k}{r}|=|\azt{k-1}{r}|$, we obtain \eqref{77} by substituting \eqref{alfrrr}, \eqref{eq:alf:sum},  and \eqref{asin} into \eqref{sar}.

\subsection{Proof of \eqref{asin}} \label{sss}
For every $0\leq m,n\leq L-1$ and $1 \leq k\leq K-1$, we have
\begin{IEEEeqnarray}{rCl}
\left|\expe\rlbo{\left.\ant{k+1}{m}\ants{k+1}{n}\right|\az{0}}\right|
&=&\left|\expe\rlbo{\expe_{\nzt{k}{m}}\rlbo{\ant{k+1}{m}}\expe_{\nzt{k}{n}}\rlbo{\left.\ants{k+1}{n}}\,\right|\az{0}}\right|\\	&\leq&\rlp{\max\limits_{\alt{k}{m}}\rlao{\left|\expe_{\nzt{k}{m}}\rlbo{\ant{k+1}{m}}\right|}}\rlp{\max\limits_{\alt{k}{m}}\rlao{\left|\expe_{\nzt{k}{n}}\rlbo{\ants{k+1}{n}}\right|}}.\label{hata}
\end{IEEEeqnarray}
The last step follows because 
\begin{IEEEeqnarray}{rCl}
	\ant{k+1}{m}=\rlp{\ant{k}{m}+\nzt{k}{m}}e^{j\gamma\dz|\alt{k}{m}+\nzt{k}{m}|^2}.
\end{IEEEeqnarray}
Using  triangle inequality, we get that
\begin{IEEEeqnarray}{rCl}
\max\limits_{\alt{k}{m}}\rlao{\left|\expe_{\nzt{k}{m}}\rlbo{\ant{k+1}{m}}\right|}
&\leq& \max\limits_{\alt{k}{m}}\rlao{|\alt{k}{m}|\cdot\left|\expe_{\nzt{k}{m}}\rlbo{e^{j\dz\gamma\rlp{|\alt{k}{m}
				+\nzt{k}{m}|^2}}}\right|}\nonumber\\
&& +\max\limits_{\alt{k}{m}}\rlao{\left|\expe_{\nzt{k}{m}}\rlbo{\nzt{k}{m}e^{j\dz\gamma\rlp{|\alt{k}{m}+\nzt{k}{m}|^2}}}\right|}\label{sor}.
\end{IEEEeqnarray}
We first compute the first term on the RHS of \eqref{sor}. Note that given $\alt{k}{m}$,  the random variable 
\begin{align}
\daf=\frac{2|\alt{k}{m}+\nzt{k}{m}|^2}{\sn^2}
\end{align}
which is proportional to the argument of the exponential term on the RHS of \eqref{sor},
follows a   noncentral chi-squared distribution with two degrees of freedom and noncentrality parameter $\lambda=2|\alt{k}{m}|^2/{\sn^2}$. Let $M_{\daf}(t)$ be the moment generating function of  $\daf$. We have \cite[Chap. 7]{papolis_PrSt}
\begin{IEEEeqnarray}{rCl}
 M_{\daf}(t)&=& \expe\rlbo{e^{t\daf}}\\
 &=&\frac{1}{1-2t}\mathrm{exp}\rlpo{\frac{\lambda t}{1-2t}}.
\end{IEEEeqnarray}
Therefore, 
\begin{IEEEeqnarray}{rCl}
|\alt{k}{m}|\cdot\left|\expe_{\nzt{k}{m}}\rlbo{e^{j\dz\gamma\rlp{|\alt{k}{m}
			+\nzt{k}{m}|^2}}}\right|
&=&|\alt{k}{m}|\left|M_{\daf}\rlpo{\frac{j\dz\gamma\sn^2}{2}}\right|\\
&=&\frac{|\alt{k}{m}|}{\sqrt{1+\sigma_n^4\gamma^2\dz^2}}\mathrm{exp}\rlpo{-\frac{\sigma_n^2\gamma^2\dz^2}{1+\sigma_n^4\gamma^2\dz^2}|\alt{k}{m}|^2}.\label{int_calcf2}
\end{IEEEeqnarray}
Using in \eqref{int_calcf2} that
\be
\max\limits_{x>0}\rla{xe^{-ax^2}}=\frac{1}{\sqrt{2ea}},\ \  a>0\label{emax}
\ee
we obtain 
\begin{IEEEeqnarray}{c}
\max\limits_{\alt{k}{m}}\rlao{|\alt{k}{m}|\cdot\left|\expe_{\nzt{k}{m}}\rlbo{e^{j\dz\gamma\rlp{|\alt{k}{m}
				+\nzt{k}{m}|^2}}}\right|}
=\frac{1}{\sqrt{2e}\sigma_n\gamma\dz}\label{pachol}.
\end{IEEEeqnarray}

The evaluation of the second therm on the RHS of \eqref{sor}  requires same care. Indeed, although the phase of $\nzt{k}{m}$ is uniformly distributed, the phase of $\nzt{k}{m}\exp\rlpo{j\dz\gamma\rlp{|\alt{k}{m}+\nzt{k}{m}|^2}}$ is not uniform since $|\alt{k}{m}+\nzt{k}{m}|^2$ depends on  $\ang{\nzt{k}{m}}$. 
To evaluate this term, we first note that
\begin{IEEEeqnarray}{rCl}
	\left|\expe_{\nzt{k}{m}}\rlbo{\nzt{k}{m}e^{j\dz\gamma\rlp{|\alt{k}{m}+\nzt{k}{m}|^2}}}\right|
	&=&\left|\expe_{\nzt{k}{m}}\rlbo{\nzt{k}{m}e^{j\dz\gamma\rlp{|\nzt{k}{m}|^2+2\red\rlpo{\alt{k}{m}\nzt{k}{m}^*}}}}\right|.\label{pari1}
\end{IEEEeqnarray}
We then separate the real and imaginary parts of $\nzt{k}{m}$ on the RHS of \eqref{pari1} to obtain
\begin{IEEEeqnarray}{rCl}
	\IEEEeqnarraymulticol{3}{l}{\expe_{\nzt{k}{m}}\rlbo{\nzt{k}{m}e^{j\dz\gamma\rlp{|\nzt{k}{m}|^2+2\red\rlpo{\alt{k}{m}\nzt{k}{m}^*}}}}}\nonumber\\	&=&\expe_{\red\rlpo{\nzt{k}{m}}}\rlbo{\red\rlpo{\nzt{k}{m}}e^{j\gamma \dz \rlp{\red\rlp{\nzt{k}{m}}^2+2\red\rlp{\alt{k}{m}}\red\rlp{\nzt{k}{m}}}}}\nonumber\\
	&&\times \expe_{\ima\rlp{\nzt{k}{m}}}\rlbo{e^{j\gamma \dz \rlp{\ima\rlp{\nzt{k}{m}}^2+2\ima\rlp{\alt{k}{m}}\ima\rlp{\nzt{k}{m}}}}}\nonumber\\
	&& +j\expe_{\red\rlpo{\nzt{k}{m}}}\rlbo{e^{j\gamma \dz \rlp{\red\rlp{\nzt{k}{m}}^2+2\red\rlp{\alt{k}{m}}\red\rlp{\nzt{k}{m}}}}}\nonumber\\
	&&\times \expe_{\ima\rlp{\nzt{k}{m}}}\rlbo{\ima\rlpo{\nzt{k}{m}}e^{j\gamma \dz \rlp{\ima\rlp{\nzt{k}{m}}^2+2\ima\rlp{\alt{k}{m}}\ima\rlp{\nzt{k}{m}}}}}.\label{parigoli}
\end{IEEEeqnarray}
One can calculate the expectations on the RHS of \eqref{parigoli} in closed form by writing them as integrals involving the Gaussian probability density function and by using  
 the following equalities:
\be
\int\limits_{-\infty}^{\infty}xe^{ax^2+bx}\di x=\frac{b\sqrt{\pi}}{2(-a)^{3/2}}e^{\frac{-b^2}{4a}} \ , \ \ \ \red(a)<0\label{eqmax}
\ee 
and 
\be
\int\limits_{-\infty}^{\infty}e^{ax^2+bx}\di x=e^{\frac{-b^2}{4a}}\sqrt{\frac{\pi}{-a}} \ , \ \ \ \red(a)<0.
\ee
Through these steps, one gets
\begin{IEEEeqnarray}{c}
	\expe_{\nzt{k}{m}}\rlbo{\nzt{k}{m}e^{j\dz\gamma\rlp{|\nzt{k}{m}|^2+2\red\rlpo{\alt{k}{m}\nzt{k}{m}^*}}}}
	=\frac{j\sn^2\gamma\dz}{\rlp{1-j\sn^2\gamma\dz}^2}\alt{k}{m}\exp\rlpo{-\frac{\rlp{\sigma_n\gamma\dz|\alt{k}{m}|}^2}{1-j\sigma_n^2\gamma\dz}}.\IEEEeqnarraynumspace\label{eqf}
\end{IEEEeqnarray}
Substituting \eqref{eqf} into \eqref{pari1}, we obtain
\begin{IEEEeqnarray}{c}
\left|\expe_{\nzt{k}{m}}\rlbo{\nzt{k}{m}e^{j\dz\gamma\rlp{|\alt{k}{m}+\nzt{k}{m}|^2}}}\right|
=\frac{\sn^2\gamma\dz}{1+\sn^4\gamma^2\dz^2}|\alt{k}{m}|\exp\rlpo{-\frac{\rlp{\sigma_n\gamma\dz|\alt{k}{m}|}^2}{1+\sigma_n^4\gamma^2\dz^2}}.\label{seeextra}
\end{IEEEeqnarray}
Using \eqref{emax} in \eqref{seeextra} we conclude that
\begin{IEEEeqnarray}{c}
\max\limits_{\al{k}}\rlao{\left|\expe_{\nzt{k}{m}}\rlbo{\nzt{k}{m}e^{j\dz\gamma\rlp{|\alt{k}{m}+\nzt{k}{m}|^2}}}\right|}=\frac{\sn}{\sqrt{2e\rlp{1+\sn^4\gamma^2\dz^2}}}\label{eq511}.
\end{IEEEeqnarray}
Furthermore, substituting \eqref{pachol} and \eqref{eq511} into \eqref{sor}, we obtain 

\begin{IEEEeqnarray}{rCl}
\max\limits_{\al{k}}\rlao{|\expe_{\nzt{k}{m}}\rlbo{\ant{k+1}{m}}|}
&\leq& \frac{1}{\sigma_n\gamma\dz\sqrt{2e}}+
\frac{\sn}{\sqrt{2e\rlp{1+\sn^4\gamma^2\dz^2}}}\label{max1}\\
&\leq&\frac{K\sqrt{K}}{\gamma Z\sqrt{2e \pn}}+
\frac{\sqrt{\pn}}{\sqrt{2e K}}.\label{erag}
\end{IEEEeqnarray}
Proceeding analogously, one can show that the same bound holds for the second term on the RHS of  \eqref{hata},. Namely,
\begin{IEEEeqnarray}{rCl}
\max\limits_{\alt{k}{m}}\rlao{\left|\expe_{\nzt{k}{n}}\rlbo{\ants{k+1}{n}}\right|}
	&\leq&\frac{K\sqrt{K}}{\gamma Z\sqrt{2e \pn}}+
	\frac{\sqrt{\pn}}{\sqrt{2e K}}.\label{erag2}
\end{IEEEeqnarray}
Substituting  \eqref{erag} and \eqref{erag2} into \eqref{hata}, we obtain after simple algebraic manipulations
\begin{IEEEeqnarray}{rCl}
\left|\expe\rlbo{\left.\ant{k+1}{m}\ants{k+1}{n}\right|\az{0}}\right|
&\leq &
 \frac{K^3}{2e\gamma^2 \pn Z^2}+\frac{K}{eZ\gamma}+\frac{\pn}{2eK}.\label{cub}
\end{IEEEeqnarray}
We note that the steps just performed are not applicable to the first segment ($k=0$) of the SSF channel as there is no noise. %

\section{Proof of Theorem~\ref{th3}}\label{apD}

Comparing \eqref{alphabound} and \eqref{cbound}, one sees that to 
 establish \eqref{cbound}, it is sufficient to show that 
\be
\zeta(\po)\po^2\geq \po^2-\frac{2\Cone}{K}\po^2-PG\label{yemosalas}
\ee
or, equivalently, that
\be
1-\zeta\rlpo{\po}\leq \frac{2\Cone}{K}+\frac{G}{P}.\label{domosalas}
\ee
It follows from \eqref{zeta} that
\begin{IEEEeqnarray}{rCl}
\zeta(P)&=&|\alf{0}{0}|^{4(K-1)}\exp\rlpo{-\frac{\UB}{P{|\alf{0}{0}|^{2(K-1)}}}}\\
&\geq &|\alf{0}{0}|^{4(K-1)}\rlp{1-\frac{\UB}{P{|\alf{0}{0}|^{2(K-1)}}}}\label{z3}
\end{IEEEeqnarray}
where in the last step we used that $e^{-x}\geq 1-x$ for all  $x\in \mathbb{R}$.
In Appendix~\ref{af11}, we prove that for every $ K>~\frac{\beta_2Z\pi^2}{2\sqrt{2}\dt^2}$ we have 
\begin{align}
|\alf{0}{0}|^2
&\geq{1-\frac{\Cone}{K^2}}.\label{alf11}
\end{align}
Substituting  \eqref{alf11} into \eqref{z3}, and using that $|\alf{0}{0}|\leq 1$ (see the definition in \eqref{Alfdef}), we obtain
\begin{IEEEeqnarray}{rCl}
\zeta(P)&=&|\alf{0}{0}|^{4(K-1)}-{|\alf{0}{0}|^{2(K-1)}}\frac{\UB}{P}\\
&\geq& \rlp{1-\frac{\Cone}{K^2}}^{2K}-\frac{\UB}{\po}.\label{z32}
\end{IEEEeqnarray}
Since the function $f\rlpo{\Cone}=\rlp{1-\Cone/K^2}^{2K}$ is convex in the interval $[0  ,  K^2]$, for every ${\Cone\leq K^2}$ or equivalently every $K\geq\sqrt{\Cone}$, we have 
\be\label{106}
f\rlpo{\Cone}\geq f(0)+f'(0)\Cone.
\ee
It follows from \eqref{106} that 
\be\label{saat3onime}
\rlp{1-\frac{\Cone}{K^2}}^{2K}\geq 1-\frac{2\Cone}{K}.
\ee
Substituting \eqref{saat3onime} into \eqref{z32} we obtain
\begin{align}\label{c}
1-\zeta(P) &\leq  \frac{2\Cone}{K}+\frac{\UB}{\po}.
\end{align}

To conclude the proof, we show that $\UB\leq\UBtw$ when $ K>\frac{\beta_2 Z\pi^2}{2\sqrt{2}\dt^2}$. Comparing \eqref{Mkdef} and \eqref{semosalas}, we see that this is equivalent to showing that 
\be
\alpha\leq \frac{\Ctwo}{K}\rlp{\frac{\Ctwo}{4K}+1}\label{dodayere}
\ee
where $\alpha$ and $\Ctwo$ are defined in \eqref{alfdef} and \eqref{c2def}, respectively. To prove \eqref{dodayere}, we show  in Appendix~\ref{afrm} that if $ K>\frac{\beta_2 Z\pi^2}{2\sqrt{2}\dt^2}$ then
\begin{IEEEeqnarray}{c}
	|\alf{0}{m}|\leq
	\frac{\Ctwo}{2KL}\label{hg21q}
\end{IEEEeqnarray}
for every $1\leq m\leq L-1$. Furthermore, by the definition of $\alpha$ in \eqref{alfdef} we have
\begin{IEEEeqnarray}{rCl}
	\alpha&=&\rlp{\sum\limits_{\substack{m=0}}^{L-1}|\alf{0}{m}|}^2-\sum\limits_{\substack{m=0}}^{L-1}|\alf{0}{m}|^2\\
	&\leq& \rlp{\sum\limits_{\substack{m=0}}^{L-1}|\alf{0}{m}|}^2-|\alf{0}{0}|^2\\
		&=& \rlp{|\alf{0}{0}|+\sum\limits_{n=1}^{L-1}|\alf{0}{n}|}^2-|\alf{0}{0}|^2\label{eqref}.
\end{IEEEeqnarray}
Using \eqref{hg21q} in \eqref{eqref}, we obtain
\begin{IEEEeqnarray}{rCl}
	\alpha&\leq&\rlp{|\alf{0}{0}|+\frac{\Ctwo}{2K}}^2-|\alf{0}{0}|^2\\
	&=&\frac{\Ctwo^2}{4K^2}+|\alf{0}{0}|\frac{\Ctwo}{K}.\label{cff}
\end{IEEEeqnarray}
We obtain \eqref{dodayere} from \eqref{cff} by using that  $|\alf{0}{0}|\leq 1$.

\subsection{Proof of \eqref{alf11}}\label{af11}
Note that
\begin{IEEEeqnarray}{rCl}
\absol{\alf{0}{0}}^2
&=&\frac{1}{L^2}\left|\sum\limits_{l=0}^{L-1}e^{-j\frac{Z}{K}\dllf\left(l\right)}\right|^2\label{arr}\\
&\geq&\frac{1}{L^2}\rlp{\sum\limits_{l=0}^{L-1}\cos\rlp{\frac{Z}{K}\dllf\left(l\right)}}^2.\label{fd}
\end{IEEEeqnarray}
Using the inequality 
$\cos(x)\geq 1-x^2/2,\label{eef}$
we obtain
\begin{IEEEeqnarray}{c}
\sum\limits_{l=0}^{L-1}\cos\rlp{\frac{Z}{K}\dllf\left(l\right)}\geq \sum\limits_{l=0}^{L-1}\rlp{1-\frac{Z^2\rlp{\dllf\rlpo{l}}^2}{2K^2}}.\label{fffsd}\IEEEeqnarraynumspace
\end{IEEEeqnarray}
One can verify that for every $K$ such that 
\be \label{kg} K>\frac{Z}{\sqrt{2}}\max\limits_{l}\dllf\rlpo{l}\geq\frac{\beta_2Z\pi^2}{2\sqrt{2}\dt^2}\ee
the RHS of \eqref{fffsd} is positive and  the inequality in \eqref{fffsd} holds also when we square both sides. Thus, if \eqref{kg} holds, we have that
\begin{IEEEeqnarray}{rCl}
\absol{\alf{0}{0}}^2&\geq& \frac{1}{L^2}\rlp{L-\frac{Z^2}{2K^2}\sum\limits_{l=0}^{L-1}\rlp{\dllf\rlpo{l}}^2}^2\\
&=&\rlp{1-\frac{\Cone}{2K^2}}^2\label{yestar}\\
&\geq&1-\frac{\Cone}{K^2}\label{dostar}.
\end{IEEEeqnarray}
Here, \eqref{yestar} follows from \eqref{c1def} and in \eqref{dostar} we used that $(1-x)^2\geq 1-2x$.

\subsection{{Proof of \eqref{hg21q}}}\label{afrm}

If $m\neq 0$, we have (see \eqref{Alfdef})
\begin{align}
\alf{0}{m}&=\frac{1}{\nt}\sum\limits_{l=0}^{\nt-1}e^{-j\frac{2\pi}{\nt}lm}e^{-j\frac{Z}{K}\dllf\left(l\right)}\\
&=\frac{1}{\nt}\sum\limits_{l=0}^{\nt-1}e^{-j\frac{2\pi}{\nt}lm}\rlp{e^{-j\frac{Z}{K}\dllf\left(l\right)}-1}.
\end{align}
Using triangle inequality,
\begin{IEEEeqnarray}{rCl}
|\alf{0}{m}|
&\leq&\frac{1}{\nt}\sum\limits_{l=0}^{\nt-1}\left|e^{-j\frac{2\pi}{\nt}lm}\right|\left|{e^{-j\frac{Z}{K}\dllf\left(l\right)}-1}\right|\\
&=&\frac{1}{\nt}\sum\limits_{l=0}^{\nt-1}\sqrt{\rlp{1-\cos\rlp{\frac{Z}{K}\dllf\left(l\right)}}^2+\rlp{\sin\rlp{\frac{Z}{K}\dllf\left(l\right)}}^2}.
\end{IEEEeqnarray}
Furthermore, using that $\cos\rlpo{x}\geq1-{x^2}/{2}$ for every $x$ and $\sin\rlpo{x}\leq x$ for every $x\geq 0$, we get
\begin{IEEEeqnarray}{c}\label{hg2}
|\alf{r}{m}|\leq\frac{1}{\nt}\sum\limits_{l=0}^{\nt-1}\sqrt{{\rlp{\frac{Z}{\sqrt{2}K}\dllf\left(l\right)}^4}+\rlp{\frac{Z}{K}\dllf\left(l\right)}^2}.
\end{IEEEeqnarray}
For every $ K>\frac{\beta_2 Z\pi^2}{2\sqrt{2}\dt^2}$, we have that $\frac{Z}{\sqrt{2}K}\dllf\left(l\right)\leq 1$. Thus, \be\label{hg1}\rlp{\frac{Z}{\sqrt{2}K}\dllf\left(l\right)}^4\leq\rlp{\frac{Z}{\sqrt{2}K}\dllf\left(l\right)}^2.\ee
Let $\Ctwo $ be defined as in \eqref{c2def}. Substituting  \eqref{hg1} into \eqref{hg2}, we obtain
\begin{IEEEeqnarray}{c}
|\alf{0}{m}|\leq\frac{\Ctwo}{2KL}\label{hg2q}.
\end{IEEEeqnarray}

\section*{Acknowledgment}

The authors would like to thank L. Barletta, N. V. Irukulapati,  G. Kramer and M. I. Yousefi  for  helpful discussions and comments on an early version of this paper.

\ifCLASSOPTIONcaptionsoff
  \newpage
\fi

\end{document}